# Autocorrective interferometers for photonic integrated circuits

Matteo Cherchi
VTT – Technical Research Centre of Finland Ltd, Tietotie 3, 02150 Espoo, Finland

## ABSTRACT

Extensive literature has shown that finite impulse response (FIR) interferometers can be engineered to be insensitive under variations of different physical parameters, e.g., to ensure flat-top response and/or tolerance to fabrication errors. In this context, I will show how the Bloch sphere representation can be a very powerful design tool providing superior physical insight into the working principle of autocorrective devices like broadband 50:50 splitters or flat-top interleavers, that can be therefore designed through simple analytical formulas. I will eventually review the recent progress in practical implementation of the autocorrective designs in the micron-scale silicon photonics platform of VTT.

## 1. INTRODUCTION

This paper is a review of my work on FIR interferometric devices for photonic integrated circuits (PICs), covering twenty years of (non-continuous) developments. It all started while studying the literature of wavelength insensitive 50:50 directional couplers, when I read an interesting paper[1] about maximally flat splitters based on generalized Mach-Zehnder interferometers (MZIs). The authors had not provided any detail how they had derived the maximally flat conditions, and I was not able to find any clue, not even contacting one of them. Out of frustration, I tried to resort to my background in quantum physics and started toying with the use of the Bloch sphere[2] as a geometric representation of the MZI operation. This effort led to my first publication on the subject[3] at the very early stages of my research career in photonics. At that point, I had realized that I was not the first to propose the use of the Bloch sphere for two-path interferometric systems, as I had been preceded by a couple of prominent scholars[4–6]. On the other hand, from an extensive literature search, I had got the impression that the geometrical representation had been regarded mainly as a theoretical curiosity, and never been put into real use for practical applications. As a matter of fact, I'm not aware of any textbook dealing with the subject. With my works, I have tried to show that the Bloch sphere is not only an abstract tool for quantum scientists but a very powerful and practical tool to design classical integrated interferometric devices, providing much deeper physical insight compared to numerical approaches. This paper is part of this very effort and, in particular, an attempt to popularize the topic among PIC designers. With this goal in mind, I will try my best to explain the key concepts without mathematical formulas, relying on the geometrical representation only. For a more rigorous mathematical treatment, I refer the reader to my previous publications cited throughout the text.

## 2. THE GEOMETRIC REPRESENTATION

All the results discussed in this paper are built around the powerful geometric representation introduced in this section. PIC implementations of two-path interferometric systems are built, in practice, out of just two basic building blocks, namely power splitters – including directional couplers and multimode interferometer (MMI) splitters – and phase shifters – including imbalanced waveguide lengths or waveguide tapering[3,7]. These systems can be treated analytically based on transfer matrix, but significant physical insight can be gained when representing them on the so-called Bloch sphere[3–5] (also known as Poincaré sphere, when used for polarisation states). All possible combinations of power splitting ratio and relative phase can be mapped on the sphere as depicted in Figure 1. Therefore, any state is unequivocally determined by two angles: $\theta$ representing the phase shift and $2\alpha$ corresponding to the splitting ratio $\sin^2(\alpha)/\cos^2(\alpha)$.

The three Stokes parameters $S_1$, $S_2$, and $S_3$ correspond to the three orthogonal axes. The intersections of $S_1$ with the sphere represent the single uncoupled waveguide modes $E_1$ and $E_2$, the intersections of $S_2$ represent the eigenmodes of synchronous directional couplers, i.e., the symmetric and anti-symmetric modes $E_S$ and $E_A$, whereas the intersections of $S_3$ represent the superpositions in quadrature and anti-quadrature $E_R$ and $E_L$.

*matteo.cherchi@vtt.fi; phone +358 40 684 9040, ORCID 0000-0002-6233-4466, www.vtt.fi

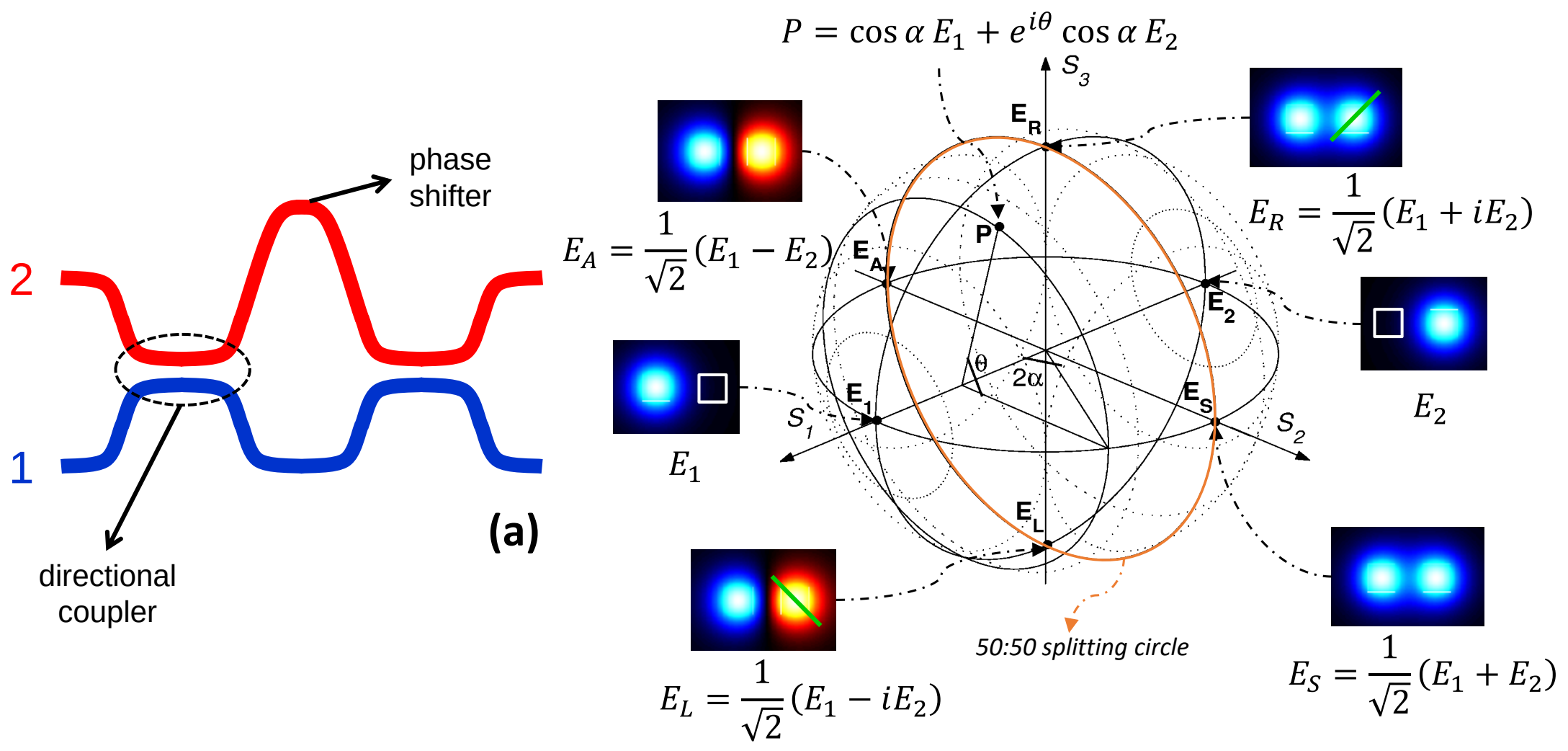

Figure 1. a) Schematic layout of a simple 2-waveguide interferometer; b) the Bloch sphere as a geometric representation of all possible states of a two-path system, showing the eigenmodes corresponding to the axes $S_1$, $S_2$ and $S_3$ and the two angles $\alpha$ and $\theta$ associated two any point P on the sphere. The angle $\alpha$ determines the power fractions $\cos^2\alpha$ and $\sin^2\alpha$ in the single waveguide modes $E_1$ and $E_2$ respectively, whereas $\theta$ is the relative phase between the two paths.

We stress that the sphere is not only a tool to show the states as static points, but also a powerful way to show their evolution under the action of power splitters and phase shifters, which operation is simply represented by rotations on the sphere. A synchronous directional coupler acts as a rotation around $S_2$, which is the axis of its eigenmodes, i.e., the symmetric and anti-symmetric modes $E_S$ and $E_A$. Similarly, a phase shifter acts as a rotation around $S_1$, which is the axis of its eigenmodes, i.e., the single waveguide modes $E_1$ and $E_2$. More in general, an asynchronous coupler acts as a rotation about the axis of its eigenstates[3], i.e., an axis laying in the equatorial plane somewhere between $S_1$ and $S_2$. As an example, we show in Figure 2 the operation of a MZI composed by two 50:50 synchronous directional couplers and a π/4 phase shift in between. The splitting ratio at the output is simply determined by the projection on the $S_1$ axis, i.e., by the angle $\alpha$. The figure also shows intuitively an important property of MZIs: the final point always ends up on the equator, meaning that the outputs are always either perfectly in phase or completely out of phase.

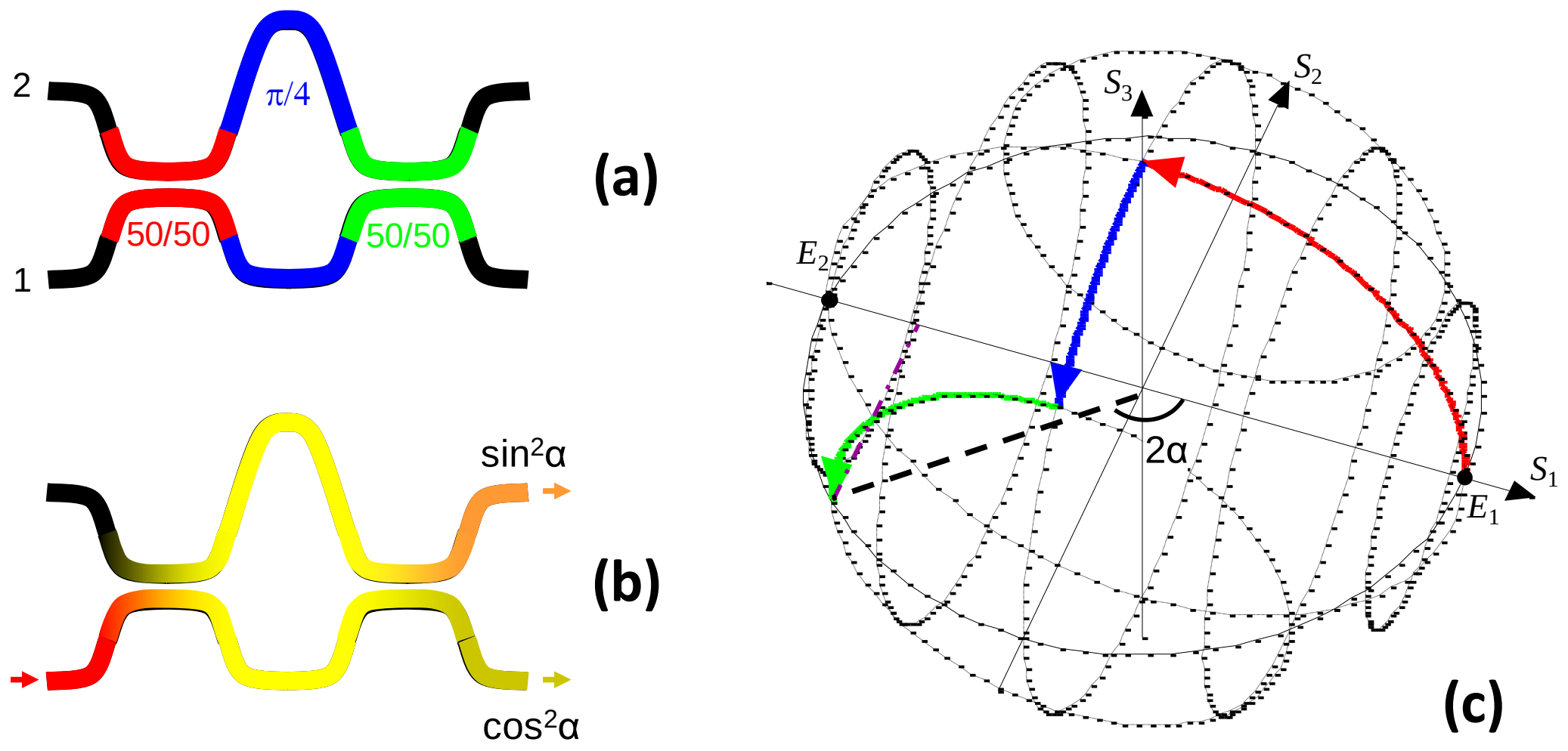

Figure 2. a) Schematic of a MZI with π/4 phase shifter; b) resulting power splitting as a function of the angle α; c) geometric representation of the MZI operation: the input state in $E_1$ is rotated by 90° around the $S_2$ axis by the first 50:50 coupler (red trajectory), therefore reaching the north pole state $E_R$. The phase shifter rotates the state by 45° around the $S_1$ axis (blue trajectory), and eventually the last 50:50 coupler rotates again by 90° around the $S_2$ axis to reach the equator circle (green trajectory).

## 3. FLAT-TOP DIRECTIONAL COUPLERS

In 1997 Little and Murphy[1] published an interesting paper about maximally flat directional couplers for different splitting ratios. The work somewhat formalized on a better theoretical ground previous experimental realisations of generalized MZIs[8,9].The aim of these works was to reduce the wavelength dependence of synchronous directional couplers. The strongest dependence corresponds to the 50:50 splitting ratio, as can be easily understood from the sinusoidal dependence of the power splitting with the coupling length (Figure 3a). This can be seen also on the Bloch sphere (Figure 3b), given that a change of the rotation angle around ±90° –due to, e.g., a departure from the nominal wavelength – corresponds to maximum change of the projection on the $S_1$ axis, which measures the power splitting.

According to Little and Murphy, the MZI with maximally flat 50:50 response turns out to be a combination of a 50:50 coupler and a 100:0 coupler (the first number representing the percentage in the cross port) with a 120° phase shifter in between them (see Figure 4), but the paper does not disclose how the result has been derived.

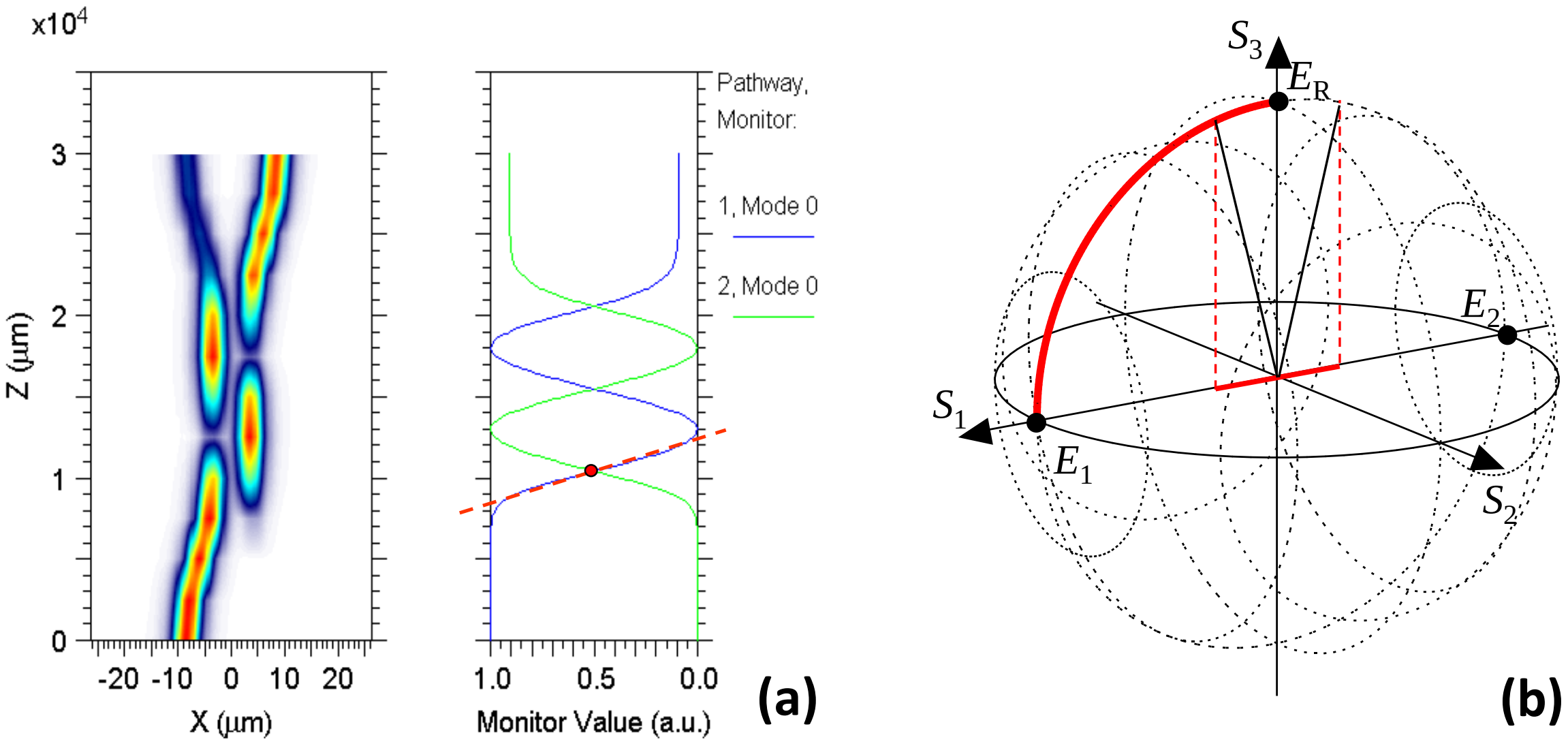

Figure 3. a) numerical simulation of a directional coupler showing the sinusoidal oscillation of the optical power between the two ports; the red dashed line highlights the point with maximum slope corresponding to 50:50 power splitting; b) Action of a 50:50 coupler on the Bloch sphere, highlighting how a variation of the rotation angle maximally translates into a variation of the splitting ratio.

### 3.1 Gaining physical insight through the geometric representation

The working principle of the flattened MZI 50:50 splitter can be easily understood using the geometrical representation. Let's first consider the configuration where the 100:0 coupler (from now on "full coupler") comes first and the 50:50 coupler (from now on "half coupler") last, which we will call the FH configuration (Figure 4a). At the nominal wavelength, the full coupler will simply switch all the power to the cross port, i.e., rotate the state by 180° from $E_1$ till the antipodal $E_2$. In this particular case, the phase shifter rotation corresponds to no changes ($E_2$ lies on $S_1$ or, in other words, it is an eigenstate of the phase shifter), and the last 50:50 directional coupler simply rotates the state by 90° to reach the south pole state $E_L$, i.e., perfect 50:50 splitting (see dashed trajectory in Figure 4a). Things become more interesting off the nominal wavelength, for example at the shorter wavelength corresponding to a reduction of 10% in the coupling angle. In this case, the full coupler will reach only till the point P, just above $E_2$ (blue continuous line in Figure 4a). This time the 120° phase shifter brings P to Q that, by construction, has a distance from the equatorial plane that is half the distance of P, but with opposite sign. This compensates exactly for the 10% reduction in the last coupling rotation, given that the reduction of a half coupler must be half that of a full coupler. Therefore, the ending point R belongs to the circle in the $S_2S_3$ plane, which is the locus of all the 50:50 states. This ensure the flat 50:50 response shown in Figure 5a. Nevertheless, R departs from the south pole $E_L$ by an angle proportional to the departure of the coupling angle from its nominal value, resulting in the linear variation of the relative phase shown in Figure 5b (dashed red line). In a similar fashion, we can appreciate the working principle of a HF configuration, where the half coupler comes first and the full coupler last (Figure 4c).

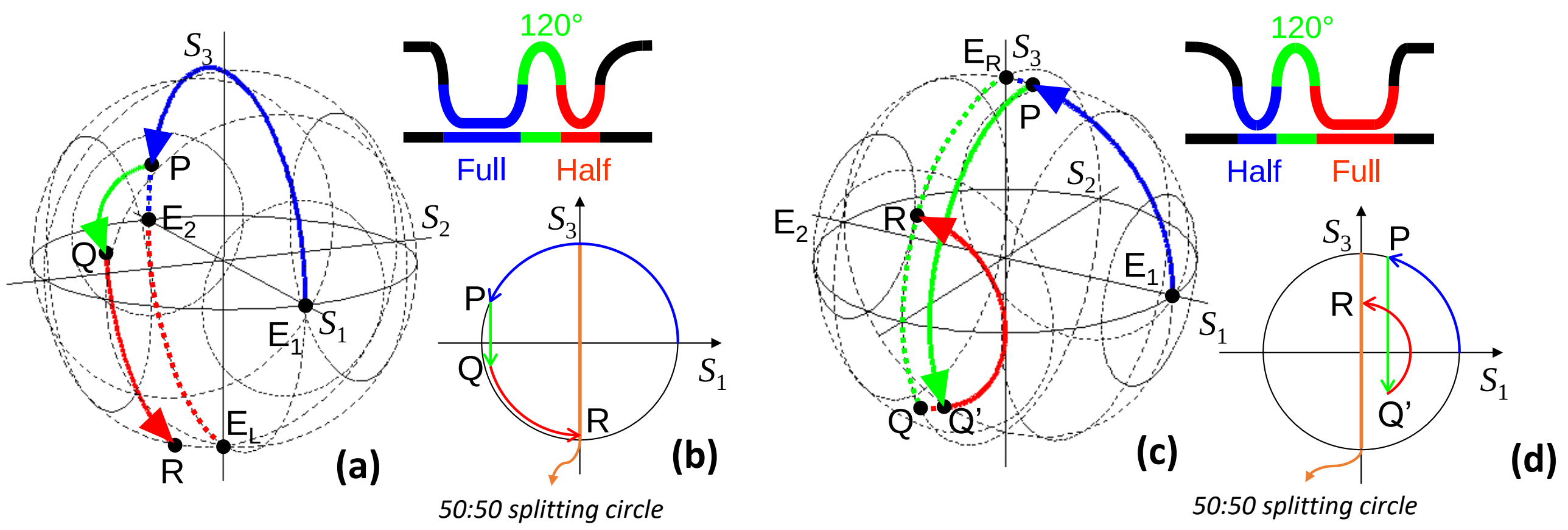

Figure 4. a) FH configuration and its trajectory on the Bloch sphere. The dashed lines correspond to the nominal case, whereas continuous lines show the autocorrective action when the coupling angles are reduced by 10%; b) projection of the same trajectory on the $S_1S_3$ plane c) and d) same as a) and b) but for the HF configuration.

In this case the nominal trajectory reaches the north pole $E_R$ first, then point Q in the southern hemisphere (with -150° phase shift) and finally point R (with 150° phase shift). All points $E_R$, R and Q belong to the 50:50 circle. At a shorter wavelength corresponding to, e.g., 10% reduction of the coupling angle, the trajectory changes, but the end points R remains substantially the same. This is because the 120° phase shift ensures that the red rotation occurs on a circle with half the radius of the blue rotation. As a result, the blue arc and the red arc are shortened exactly by the same length because, the coupling angle changes proportionally to the angle itself and, by construction, the coupling angle of the full coupler is twice the angle of the half coupler. Therefore, this configuration ensures flat response for amplitude and phase at the same time, as shown in Figure 5. In the next section, I will show how the different behaviour of the HF and FH configuration must be taken into account when designing MZI interferometers.

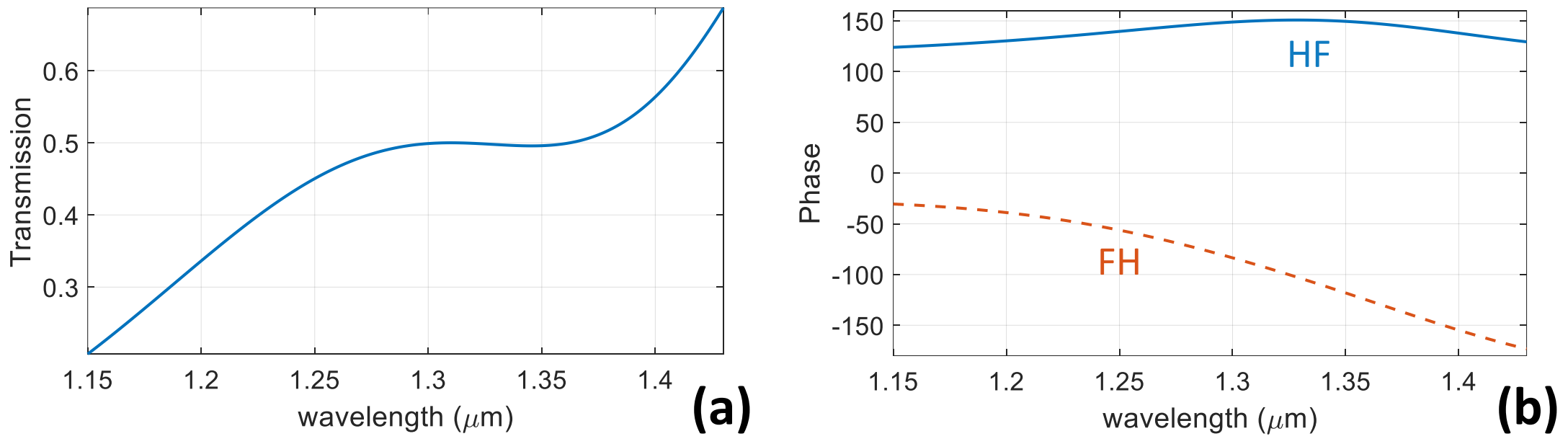

Figure 5. Example implementation of the flattened couplers a) Simulated power splitting in a glass waveguide coupler (exactly the same for both FH and HF configuration); b) simulated relative phase at the output of the coupler for both configurations; all simulation assume silica waveguides with 4.5% index contrast and square cross section with 2 µm side[10];

Notice that the rotations in Figure 4a and b assume no wavelength dependence of the phase shifter, which is reasonable for phase shifters of the lowest order (e.g., subwavelength extra length). This is confirmed by the numerical simulations in Figure 5, that take into account the actual wavelength dependence of the phase shifter.

From this example application to simple MZI couplers, it should be now clear that the Bloch sphere provides much deeper insight in the operation of two-path FIR interferometers, highlighting properties – like the phase response – that had been completely missed in previous works.

### 3.2 Application to interferometers

The flattened 50:50 splitters introduced in the above section can be used, for example, to build wavelength insensitive Mach-Zehnder interferometers. Both the FH and HF configuration will work fine, as long as they are used in a point-symmetric configuration[11], as depicted in Figure 6a and b. This can be easily understood on the Bloch sphere, considering that a point-symmetric device results in a mirror-symmetric trajectory (Figure 6c) because, by construction, rotations of couplers remain the same, whereas rotations of phase shifter have opposite sign[10]. In fact even if in the FH configuration the point R moves with wavelength on the 50:50 circle, by symmetry it moves by the same amount for both splitters. Things become more complicated when considering Michelson interferometers. A common way to realise add-drop filters is to place identical grating filters on the two arms of a balanced MZI[12]. The device still works as a MZI for the transmitted wavelengths but turns to a Michelson interferometer for the reflected wavelengths. Using the Bloch spere, it can be shown[10] that such interferometer requires a mirror symmetric configuration corresponding to a point symmetric trajectory. This means that the 50:50 couplers must end up in either one of the two poles $E_R$ or $E_L$, which projection on the $S_1S_2$ plane is the centre O of equatorial circle (Figure 7c).

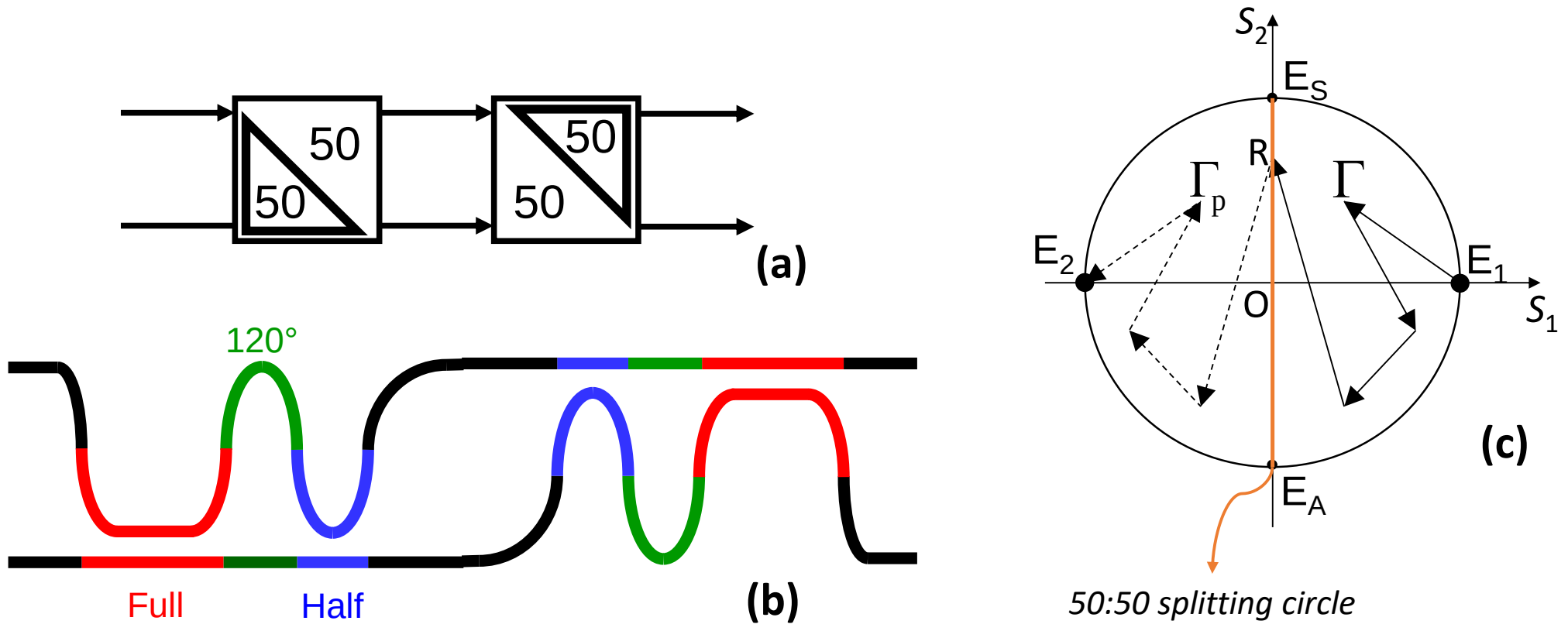

Figure 6. a) Schematic representation of a balanced MZI formed combining a generic 50:50 splitter with a point-symmetric replica of the same; b) schematics showing how to form a wavelength insensitive MZI with two FH flattened splitters in point-symmetric configuration; c) working principle of the point-symmetric configuration explained for a generic 50:50 splitter, showed through the projection of a generic trajectory on the equatorial plane $S_1S_2$; the resulting mirror-symmetric trajectory always ends up in the antipodal point $E_2$, no matter how the point R changes its position on the 50:50 circle.

It is clear that the FH configuration cannot work properly in this case, because the trajectory reaches the south pole only at the nominal wavelength, to depart significantly at different wavelengths. On the other hand, the HF configuration has a stable ending point, which is far from either one of the poles. This is not a major issue, given that we can always reach the north pole adding a fixed -60° phase shift (with negligible wavelength dependence when choosing the lowest order).

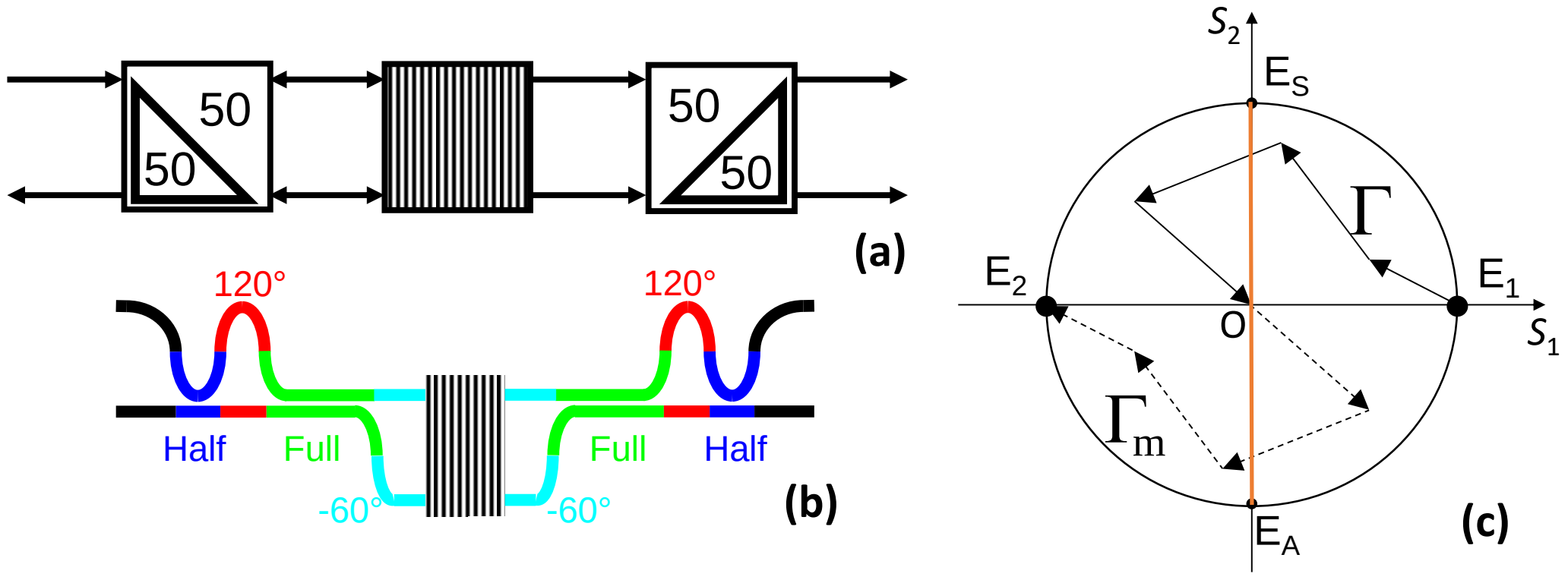

Figure 7. a) Schematic representation of a add-drop filter using a generic 50:50 splitter with a mirror-symmetric replica of the same; the grating filter reflects back some of the wavelengths b) schematics showing how to form a wavelength insensitive Michelson interferometer with two HF flattened splitters in mirror-symmetric configuration; c) working principle of the mirror-symmetric configuration explained for a generic 50:50 splitter, showed through the projection of a generic trajectory on the equatorial plane $S_1S_2$; the resulting point-symmetric trajectory must pass through the central point O, that is the projection of either one of the poles.

The full 3D trajectory of the interferometer sketched in Figure 7b is shown in Figure 8a, at the wavelength corresponding to coupling rotations 10% smaller than the nominal values. It can be seen how the phase shifters pass through the north pole (cyan trajectory) and that the ending point remains very close to the $E_2$ cross state. The simulation of a practical implementation on a glass waveguide platform confirms a broadband flat response exceeding 200 nm bandwidth at 1 dB, as shown in Figure 8b.

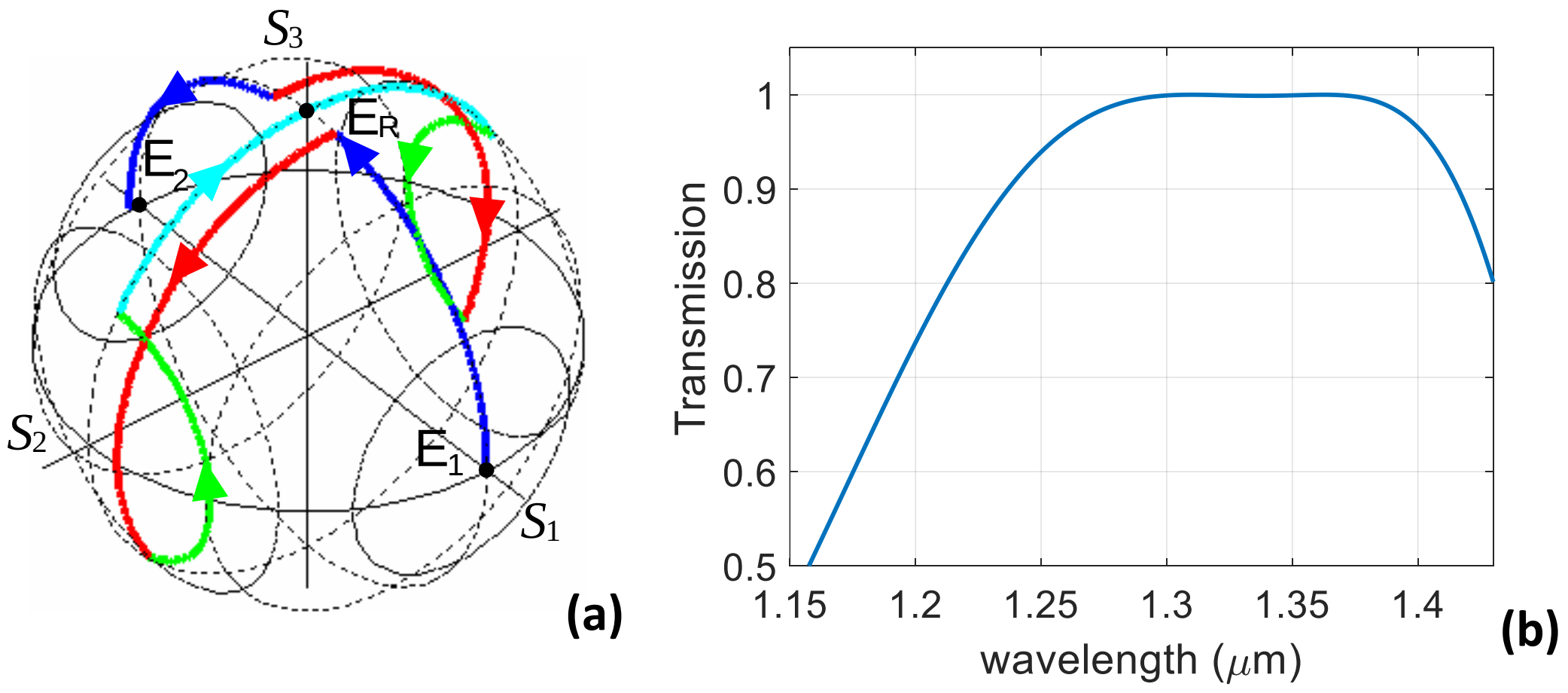

Figure 8. a) Trajectory on the Bloch sphere of the mirror symmetric interferometer in Figure 7b, following the same colour coding, when coupler rotations are 10% smaller than the nominal value; b) simulated response of the same interferometer implemented in a silica waveguide platform.

## 4. FLAT-TOP FILTERS

The Bloch sphere provides major physical insight also in the design of lattice filters and generalised lattice filters[13]. We will first show the working principle of a standard two-stage lattice filter, and how this allows to identify all possible solutions through simple analytical formulas. We will then use the geometric representation to analyse a particular class of four-stage MZI called doubly symmetric interferometers, to show how their design can be made more systematic and even extended to a broader class of configurations.

### 4.1 Two-stage generalised lattice filters

The geometrical representation can easily explain the working principle of two-stage generalised lattice filters, where the imbalance between the two arms doubles at each stage. The usual design approach is based on numerical optimization algorithms to determine the coupling coefficients of the splitters. Compared to standard lattice filters, generalized lattice filters can achieve similar performance with fewer stages. For example, a two-stage generalized lattice filter can achieve comparable flat-top band and roll-off of a standard four-stage lattice filter, like the ones analysed in §4.2.

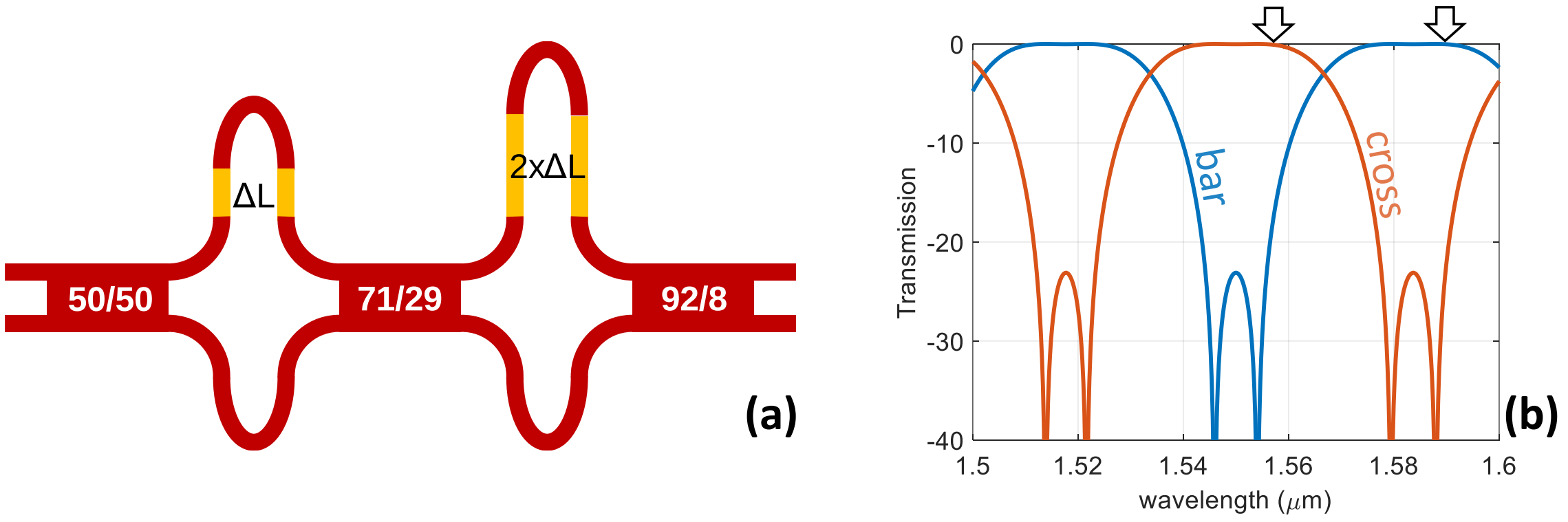

Figure 9. a) Schematic layout of the 2-stage flat-top generalised lattice filter; b) simulated spectral response of a possible physical realisation of the filter.

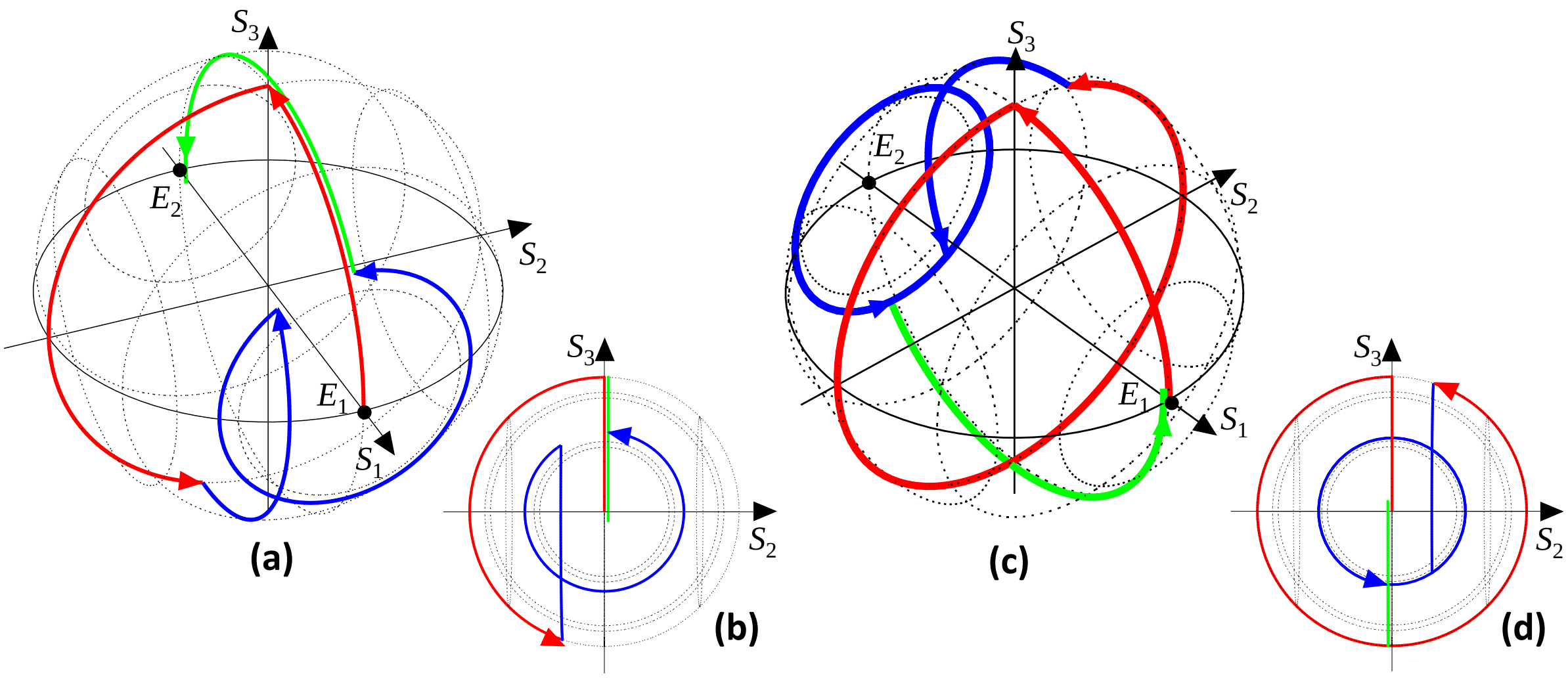

Figure 10. a) Operation of the filter for a wavelength slightly longer than the central wavelength of a cross channel, as shown by the arrow in Figure 9b, corresponding to a phase shift smaller than the nominal value; b) projection on the $S_2S_3$ plane of the same trajectory, highlighting the autocorrective behaviour of the two phase shifters; c) and d) the same for a wavelength slightly longer than the central wavelength of a bar channel.

Let's consider one of the examples reported by Madsen and Zhao[13] (page 221 of their book, Figure 4-49,), with couplers of 0.500, 0.7143, and 0.9226, and phase shifters of order +1 and +2 (but the interpretation can be easily extended to any of the other examples reported in the book). We sketch in Figure 9a a possible physical realisation of the filter, with the phase shifters implemented as arms of different physical length. The transfer matrix response of the filter is plotted in Figure 9b. Central wavelengths of transmission bands in the cross-port correspond to a $\pi+2k\pi$ ($k$ integer) phase shift in the first phase shifter, whereas for the bar port, central wavelengths correspond to $2m\pi$ ($m$ integer) phase shift in the first phase shifter. For wavelengths slightly off the central wavelengths, the phase shift will depart from multiples of $\pi$, which, in the case of simple MZIs, leads to reduced transmission and extinction ratio, i.e. to non-flat response. Instead, the 2-stage filter is designed such that the phase offset in the first stage is compensated by the offset in the second stage, resulting in autocorrective behaviour and consequent flat response. In fact, the angular offset in the second stage is, by construction twice as the angular offset in the first stage, but it occurs on a circle that has about half the radius, and in the opposite direction. This is shown in Figure 10 for the cross port and bar port respectively. The half radius is guaranteed by having chosen a 71:29 splitter, i.e. very close to 75:25 that would correspond to exactly 120° rotation. This way, the physical offsets cancel out, having same magnitude and opposite sign. Choosing a coupler rotation slightly smaller than 120° helps ensuring even wider flat-top operation. This degree of freedom can be fine-tuned to find a trade-off with roll-off, extinction ratio, and in-band ripple. In a previous paper[14], I have shown how the geometric representation allows to identify all possible solutions with simple analytic formulas. One advantage of deriving analytic formulas, compared to the traditional numerical optimisation algorithms, is that also the wavelength dependence of the couplers can be taken into account, to design filters covering very broad wavelength ranges[15,16].

We notice that the working principle of these two-stage lattice filters somewhat resembles the one of flat-top interferometric splitters with 120° phase shifters presented in §3.1, but in a complementary way, given that, in that case, the wavelength dependent elements were the couplers, not the phase shifters.

## 4.2 Four-stage lattice filters

Unlike the generalized lattice filters of the previous section, standard lattice filters[13] (also known as Fourier filters) have the same imbalance in all stages. In 1996 Jinguji et al. introduced the idea of doubly-symmetric four-stage MZIs[11], exploiting the symmetries of the lattice as a tool to ensure flat-top operation. In fact, the symmetries can be used to achieve auto-compensation of the phase errors that occur when the wavelength departs from the central wavelength of a given channel, i.e., when the relative phase between the two arms departs from multiples of $\pi$. The main issue with this approach is the limited adjustability of in-band ripple, roll-off and extinction ratio. Furthermore, it requires double the number of stages compared to the generalized lattice filters of §4.1, which in general implies larger footprint and higher losses. In the original paper, the coupling coefficients of the splitters had been derived using a numerical optimization approach.

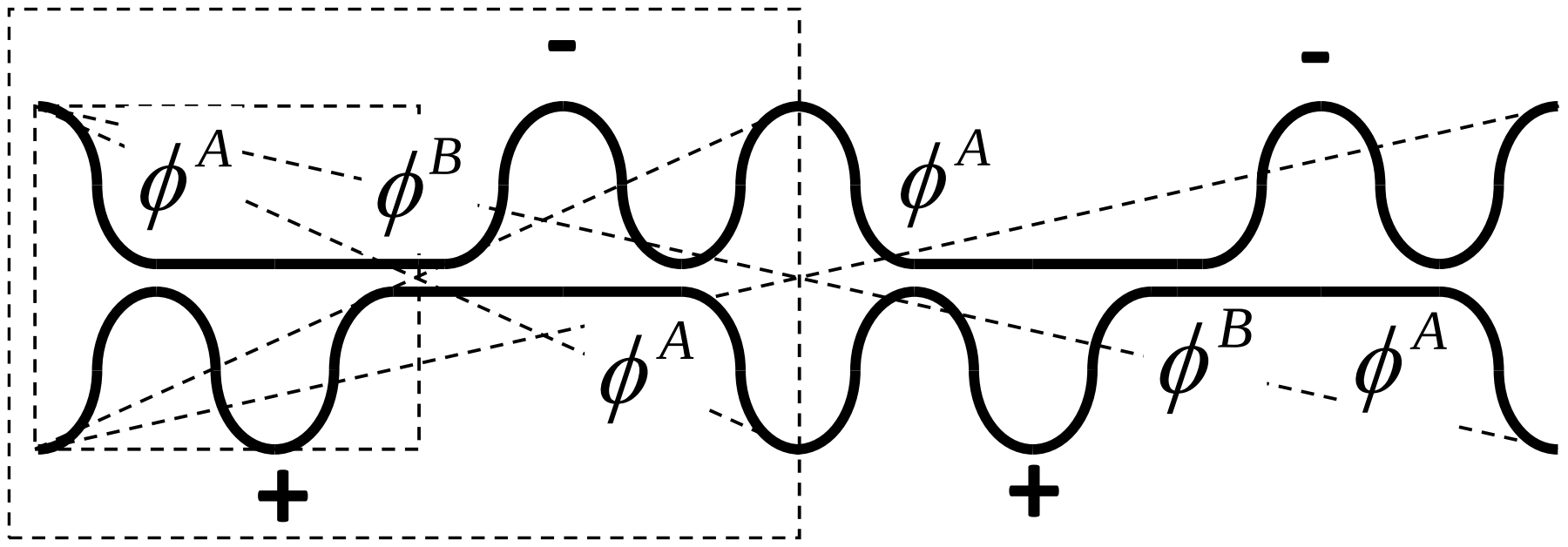

Figure 11. Schematic of a double point-symmetric configuration, showing the coupling angles $\phi^A$ and $\phi^B$ of the couplers. The starting simple MZI on the left is used to create the doubly symmetric configuration in two steps.

Using the geometrical representation, I was able to demonstrate[15] that the class of suitable symmetries and working devices can be extended, by using a simple analytic formulas, providing much deeper physical insight compared to the numerical approach of the original paper. The original idea by Jinguji and colleagues was to start from a basic building block in the form of a generalised MZI, composed by two couplers, with coupling angles $\phi^A$ and $\phi^B$, and a phase shifter in between. They first cascade to the building block a point-symmetric replica, resulting in a two-stage lattice filter with phase shifters of opposite sign. Eventually, they cascade a point symmetric replica of the resulting two-stage filter to obtain a doubly-point-symmetric four-stage lattice filter (see Figure 11) with a $+-+-$ phase shifter configuration. In the original paper, the angles $\phi^A$ and $\phi^B$ are the unknown quantities determined through a numerical optimisation algorithm. Using the geometrical representation I have been able to explain the working principle of the devices and to derive analytically all possible solutions for the coupling coefficient combinations[15]. Furthermore, I have shown that also the class of singly-point-symmetric devices with $++--$ phase shifter configuration leads to flat-top solutions. Indeed, the solution with shortest possible directional couplers relies on that alternative configuration.

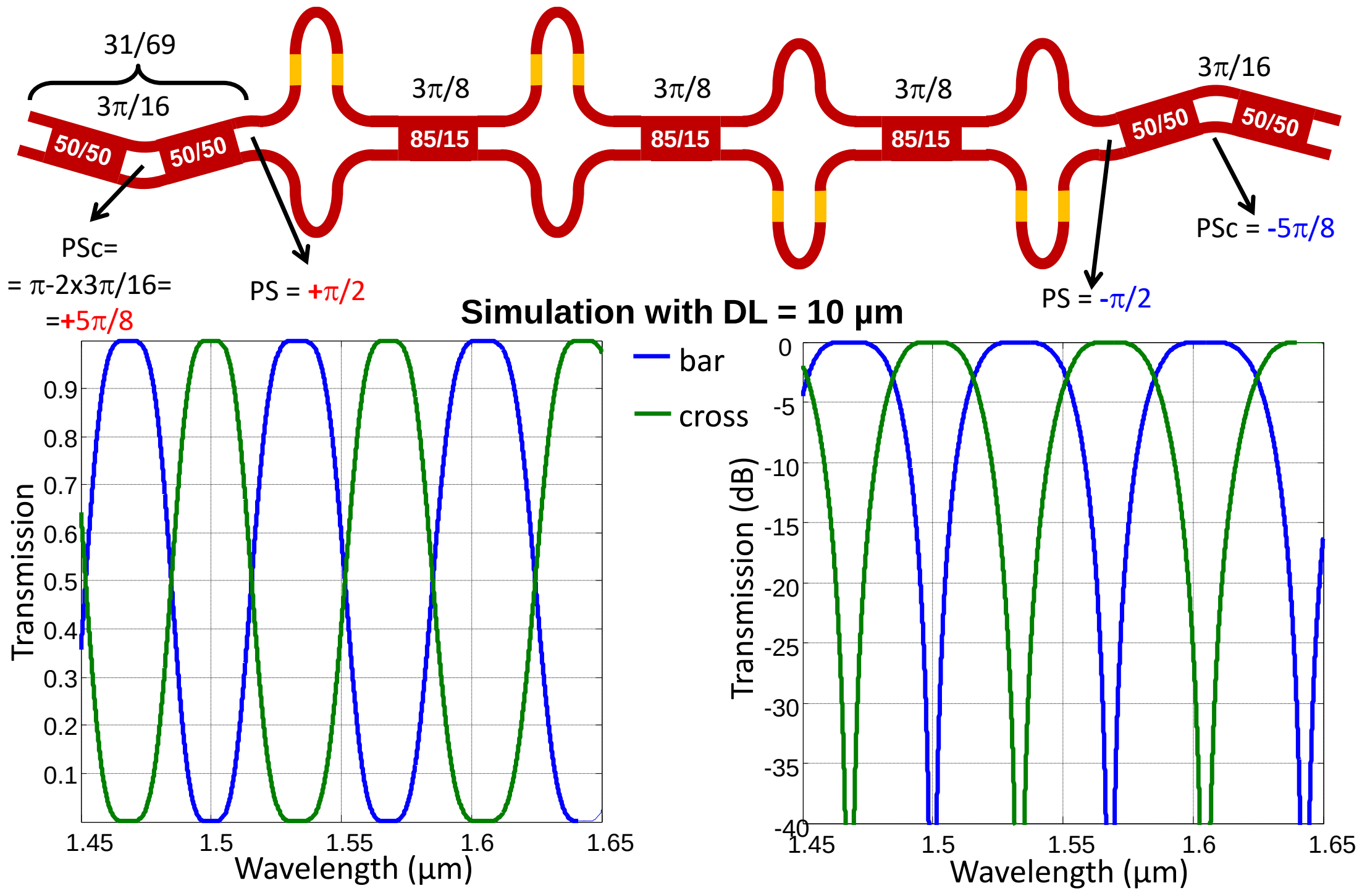

Figure 12. The proposed flat-top interleaver, showing the coupling angles of the different couplers, and the simulated flat-top response in linear scale and in dB scale.

In a more recent work[17], I have also shown how the singly-point-symmetric configuration is particularly suitable to realise flat-top interleavers using MMI splitters instead of directional couplers. An example is shown in Figure 12 together with the simulated flat-top response. Flat response for both ports can be achieved only using the $++--$ phase shifter configuration. In that particular realisation, we resorted to an MZI with bent connecting waveguides to achieve the 31:69 splitting ratio required for the first and last splitters, in combination with 90° phase shifters to mimic the phase output of standard couplers. The operation on the Bloch sphere of the of the full interleaver is shown in Figure 13, for wavelengths slightly shorter than a nominal cross port wavelength (corresponding to a phase shift slightly larger than zero) and a nominal bar wavelength (corresponding to a phase shift slightly larger than 180°). We stress that the design of such a complex interferometric device would have not been possible without the geometrical representation.

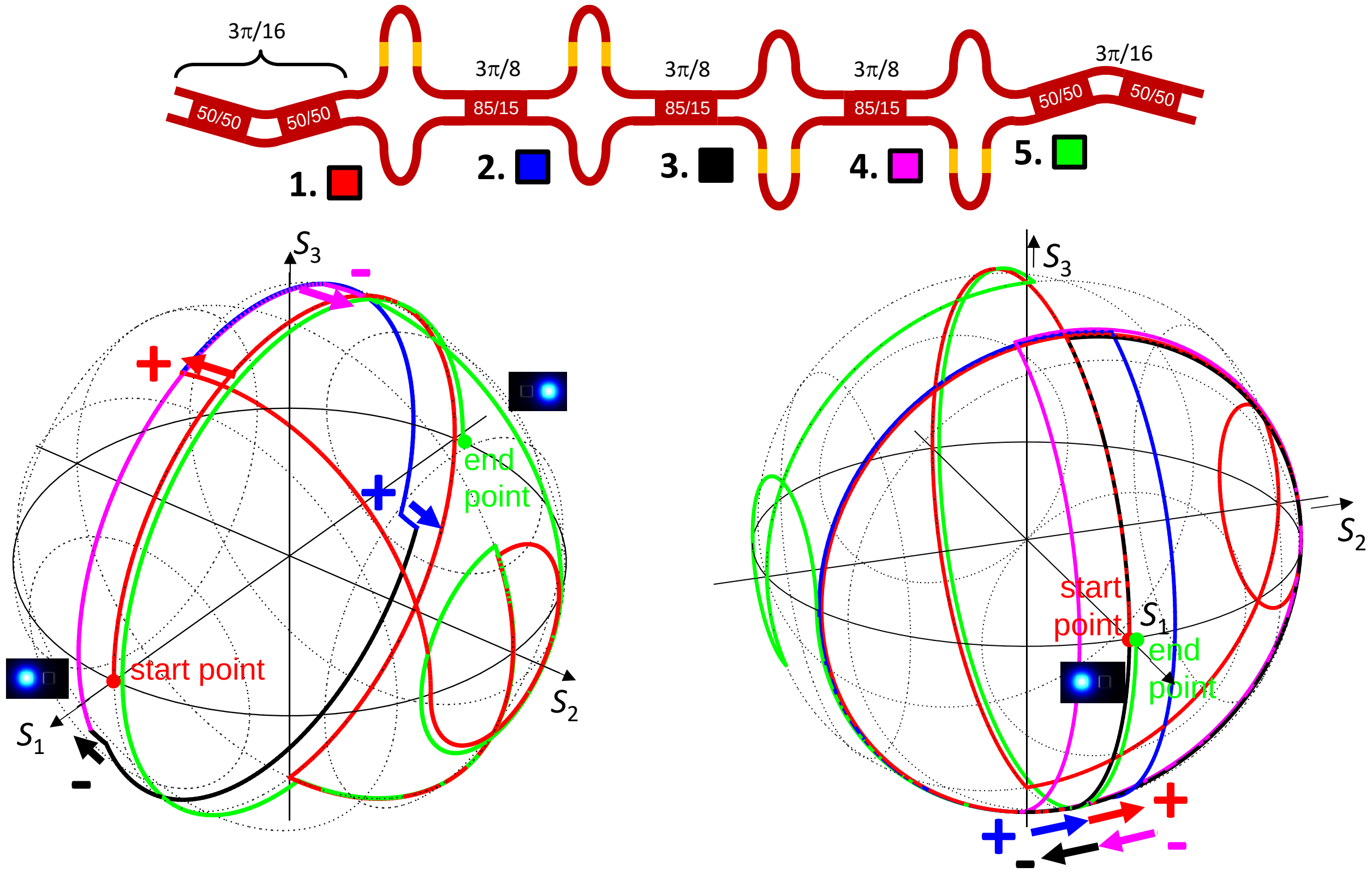

Figure 13. Autocorrective action of the four phase shifters for wavelengths slightly shorter than the in-phase (left) and in-anti-phase (right) condition. The rotations of the first four couplers and phase shifters are colour coded in pairs (red, blue, black and pink) whereas the rotations of the fifth coupler are green. Furthermore, the coloured arrows highlight how, on the left, the error in the last (pink) and third (black) phase shifters compensate for the departure of the first (red) and second (blue) phase shifter respectively whereas, on the right, the error in the second (blue) and last (pink) phase shifters compensate for the departure of the first (red) and the third (black) phase shifter respectively.

## 5. DEMONSTRATIONS ON A THICK SOI PLATFORM

We have implemented some of the interferometric filters presented in the previous section on the thick SOI platform developed by VTT. I will present the main results in the following sections, after a brief introduction about the platform.

### 5.1 VTT thick SOI platform

VTT platform is one of the few silicon photonics platforms based on micron-scale thick device layer[18,19], and the only one providing open-access[20], also through multi-project wafer (MPW) runs. We show in Figure 14 an overview of the platform, including the main waveguide types and cross-sections. The platform has a unique combination of tight bends[21–23] and relatively large modes, enabling dense integration with low propagation losses (≈ 0.1 dB/cm, down to 4 dB/cm demonstrated[24]) and low coupling losses (≈0.5 dB to tapered fibres) on a very broad band.

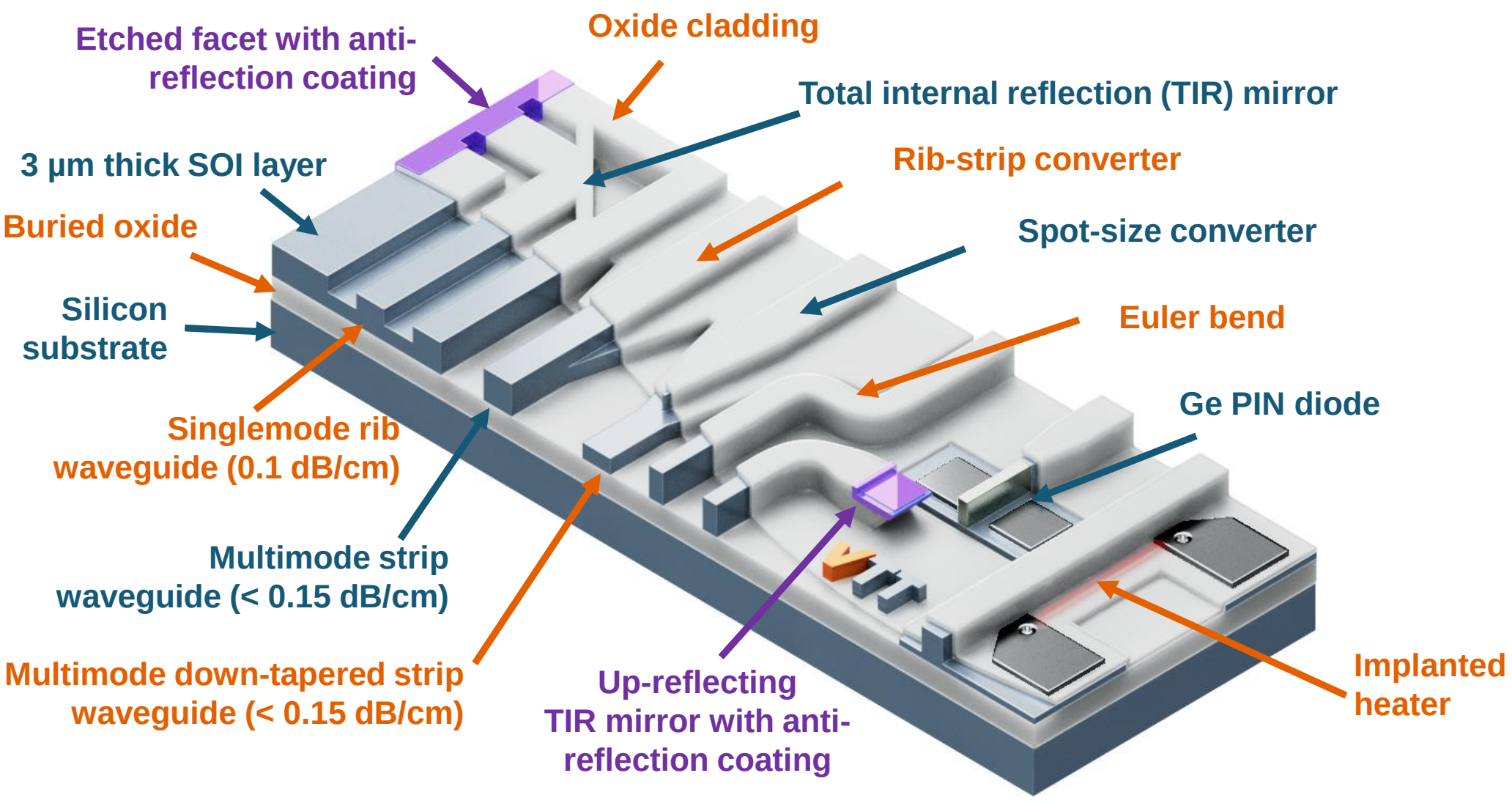

Figure 14. Sketch of the main building blocks available on the thick-SOI platform. Typical thickness of the device layer is 3 µm, whereas the buried oxide (BOX) thickness can vary from 400 nm to 3 µm.

Waveguides can be coupled with equally low-loss in-plane, through facets fabricated at wafer scale, or out-of-plane, through wet-etched 45° up-reflecting mirrors based on total internal reflection (TIR). In both cases, an anti-reflection coating is deposited at wafer scale.

Rib waveguides in silicon layers of any desired thickness can be designed to be singlemode[25] for both polarisations at the same time. On the other hand, they are not suitable to achieve tight bends, unless TIR mirrors are used (Figure 14), with the drawback of non-negligible losses[23] (0.1d dB to 0.3 dB per turn). Instead, strip waveguides are highly multimode, but they allow tight bends. In order to achieve high integration densities, we tend to prefer strip waveguides in our circuits, even though special care must be taken in order to ensure effective singlemode operation. First of all, input and output couplings happen through singlemode rib waveguides. Light is then coupled from the rib waveguide to the strip waveguide and vice versa through adiabatic rib-to-strip converters[26], with negligible loss (≈ 0.02 dB), which excites only the fundamental mode of the strip waveguide. Also the tightest Euler bends[21] are designed to ensure that most of the light remains in the fundamental mode, again resulting in very low-losses (≈ 0.02 dB).

Another important advantage of the platform, is that all waveguides support TE and TM modes with comparable confinement and losses, meaning that both polarizations are supported and that polarization insensitive devices can be achieved[18]. Furthermore, the tight confinement of the mode inside silicon, results in reduced sensitivity to fabrication variations[27] compared to submicron waveguides, supporting high production yields[19].

Power splitters in the platform can be realised either based on rib directional couplers or on strip MMI splitters. The main limitations of the directional couplers are the large footprint and the sensitivity to etch depth variations at wafer scale (typically ±5%). The main limitation of standard MMIs is that they cannot cover a continuum of splitting ratios. Nevertheless, this problem can be solved with more advanced tapered designs[14,28,29]. MMIs can be also designed to work for both polarisation at the same time, at the cost of larger footprint.

### 5.2 Four-stage lattice filters

We have implemented on the thick SOI platform the four-stage lattice presented in §4.2[17]. A first layout exactly matched the linear cascade shown in Figure 12. In a second layout, we instead warped the structure in the compact spiral shape shown in Figure 15. In both cases, we achieved the targeted flat-top response, despite some nonideality. We stress that the devices have no heater for fine tuning, meaning that all four phase shifters work perfectly in phase. This is a proof of the robustness of the platform to fabrication errors. Nevertheless, the limited extinction ratio of the measured results is mainly due to non-ideal operation of the 31:69 MZI couplers, whose inner and outer phase shifters are particularly sensitive to linewidth errors in the fabrication process. For this reason, in the subsequent designs we have moved to tapered MMIs, which tolerate linewidth changes much better.

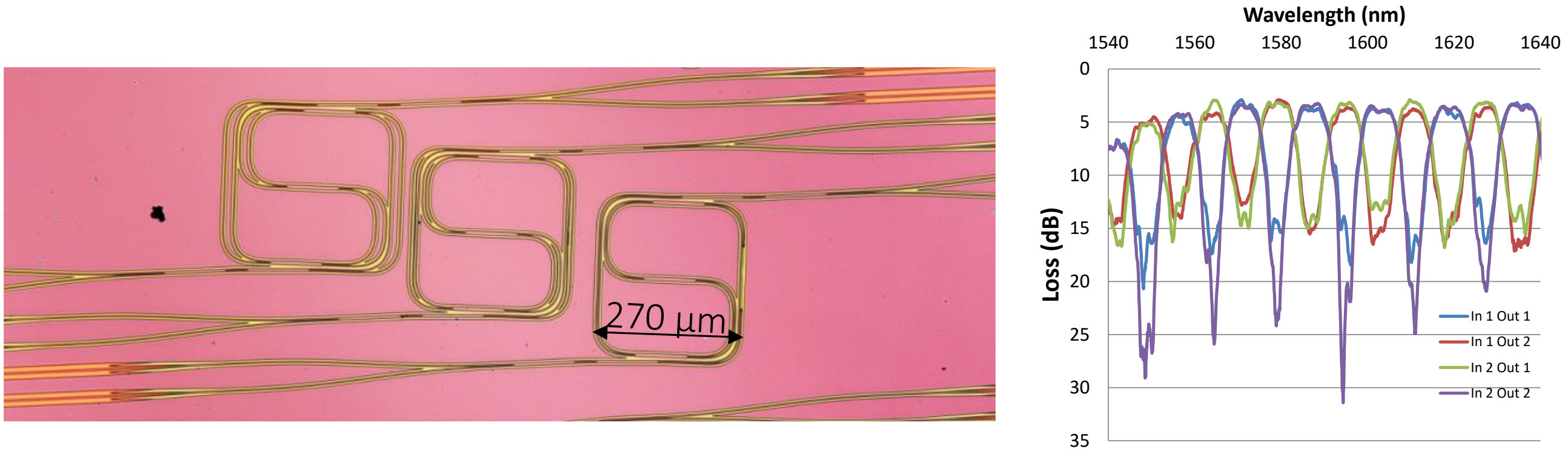

Figure 15. Micrograph of the three fabricated four-stage lattice filters with a spiral layout (left) and measured spectral response of a fabricated device (right).

## 5.3 Two-stage generalised lattice filters

We have demonstrated the two-stage lattice filters introduced in §4.1 for two different applications. In a first realisation we have designed and fabricated flat-top interleavers with different free spectral range (FSR) values using strip waveguides and tapered MMI couplers. In a second realisation, we have been able to design an ultra-broadband diplexer for a bidirectional optical sub-assembly (BOSA), based on rib waveguides and directional couplers.

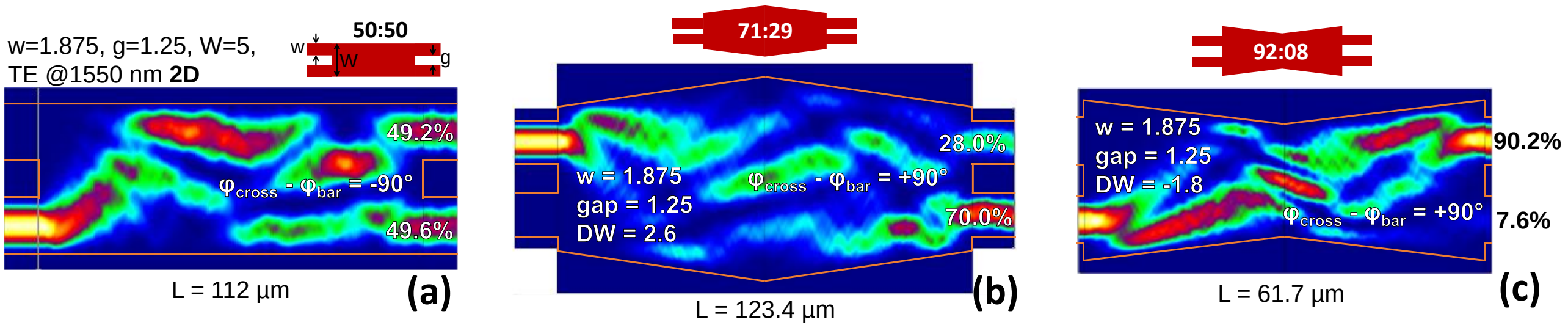

Figure 16. 2D simulations (eigenmode expansion) of the three MMI splitters used for two-stage flat-top filters; the pictures show the physical dimensions, the actual simulated transmissions, the normalised splitting ratios, and the phase shifts between the output ports.

The first realization is almost identical to the one in Figure 9, for which we have designed the MMI splitters in Figure 16 to achieve the required splitting ratios. A small but important difference is that the output phase of the 50:50 MMI (Figure 16a) has opposite sign compared to a standard lowest order directional coupler with 90° rotation on the Bloch sphere, and it is therefore equivalent to a 270° coupling rotation, reaching till the south pole state $E_L$ (as can be seen, for example, in the trajectories of Figure 13). The autocorrective operation is basically the same, with the only difference that, in this new implementation based on MMIs, wavelengths coupled to the cross-port and the bar port correspond to phase shifts which are even multiples of π and odd multiples of π respectively[14], i.e., exactly the opposite of what shown in Figure 10.

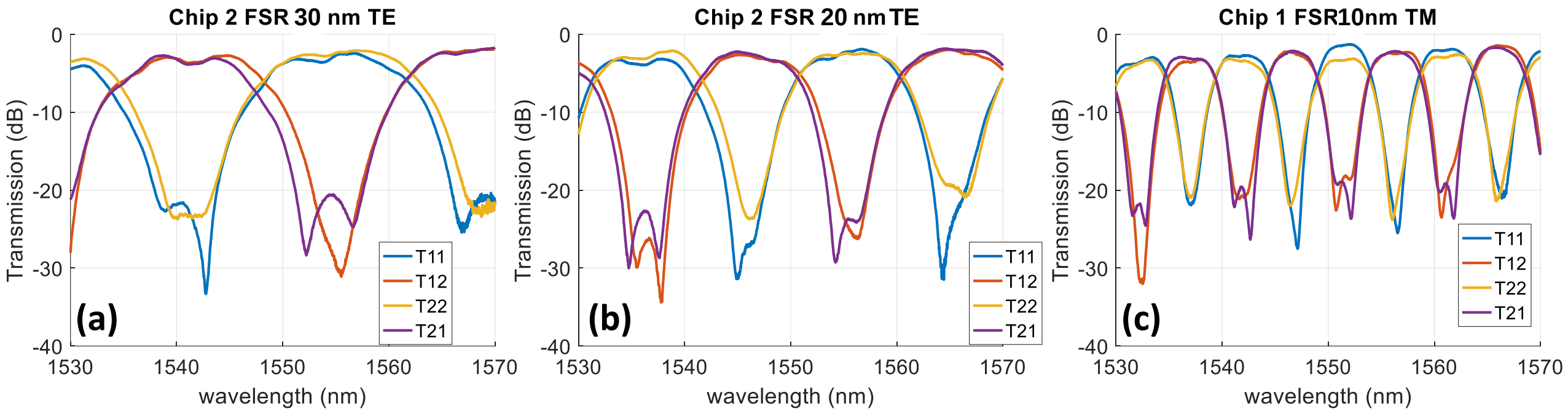

Figure 17. Measured spectral response of three different two-stage interleavers with different FSR.

We show in Figure 17 the measured spectral response of some of the fabricated devices, corresponding to different FSRs and different polarization states. We also measured the spectral response of the single MMIs, confirming that they match the designed splitting ratios over a broad wavelength range and with high tolerance to fabrication variations. We stress that, also in this case, the devices were fully passive, without any heater to fine tune the two phase shifters, showing again high tolerance of the thick SOI platform to fabrication variations.

The second implementation is an ultra-broadband filter to diplex 1310 nm and 1490 nm wavelengths. In this case, we have resorted to rib waveguides and directional couplers, given that the wavelength range of operation of MMI splitters is typically limited to about 100 nm. The design is conceptually similar to the example reported by Madsen and Zhao[13] (page 221 of their book, Figure 4-50,), with couplers of 0.500, 0.2857 and 0.0774, and phase shifters of order +1 and -2 (see Figure 18a). The problem is that their design approach assumes wavelength independent couplers, which would result in a completely spoiled spectral response over the spectral region of our interest (as highlighted in Figure 4-50 of the book). Instead, thanks to the geometrical representation, we have derived simple analytical formulas[14,15] that allow to account for the wavelength dependence of the splitters, and to achieve flat top operation over such a broad band of operation. The simulated spectral response of the designed diplexer is shown in Figure 18. It can be seen that it works for both polarisations and with negligible polarisation dependent loss. We have used the diplexer to design and fabricate a PIC-based BOSA[16], where two diplexers had been cascaded to achieve higher isolation. The overall measured excess losses were below 1.5 dB, meaning less than 0.8 dB per stage.

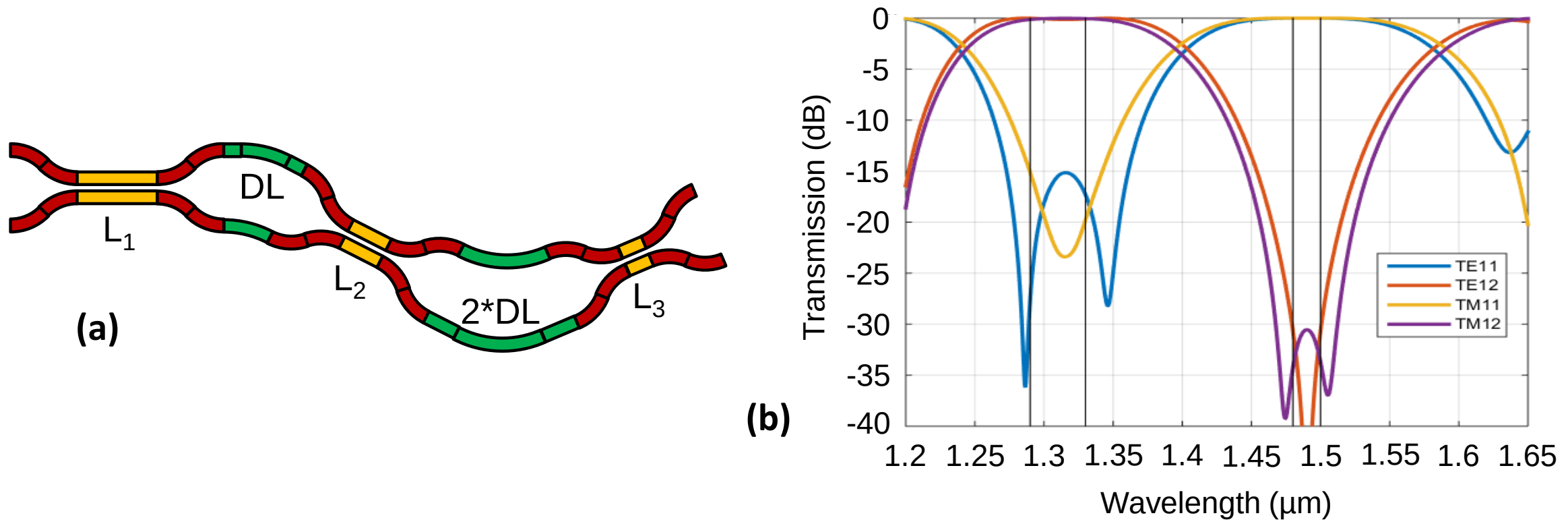

Figure 18. a) Sketch of the diplexer layout b) simulated spectral response of the designed diplexer, taking into account realistic models for the directional couplers coming from both simulations and measurements of the basic building blocks.

## 6. CONCLUSIONS

I have tried my best to show some examples how to apply the very abstract tool called Bloch sphere to very applied problems in integrated optics. The geometrical representation is very popular in quantum physics, as well as in classical optics when dealing with light polarisation, under the name of Poincaré sphere (and for good historical reasons, given that Poincaré was the first to introduce it at the end of the 19$^{th}$ century[30]). Unfortunately, its application to classical two-path interferometers has never gained popularity, despite several attempts[4–6,31–36] to introduce it to the photonics community. One of the possible reasons is that there have been very few publications showing a competitive advantage of the geometrical approach compared to more traditional mathematical and numerical approaches. In other words, it has never been shown to be useful enough to be worth spending some time to learn it. Through the collection of results in this review paper, I hope I was able to start building a critical mass of results substantiating the unmatched physical insight and design capabilities of the Bloch sphere for PIC design. My first wish is to see the geometrical approach put into systematic use by PIC designers to solve practical problems, hopefully well beyond the examples shown in this work, but also to see the topic covered in textbooks and academic courses in the near future.

A second wish is to reach those scientific communities that use the Bloch sphere as everyday tool to deal with so called two-level systems. Even though the present paper is focused on wavelength flattened optical filters, the autocorrective mechanisms presented throughout the paper are potentially applicable to any two-level system, thanks to the universality of the Bloch representation. For example, it would be interesting to explore whether similar mechanisms could help to implement robust quantum devices, even though I can already see some hurdles in that direction. In fact, a general learning from all examples in this paper is that, whenever an element (e.g. a coupler or a phase shifter) suffers from errors, it is possible to achieve an autocorrective configuration by ensuring that errors from multiple elements of the same kind cancel each other. This is true under the assumption that the relative errors of all those multiple rotations are exactly the same (i.e. they are perfectly correlated), which typically is not the case in quantum systems.

## ACKNOWLEDGEMNTS

I acknowledge support by the Academy of Finland Flagship Programme, Photonics Research and Innovation (PREIN), decision number 320168, and by the Business Finland project PICAP, decision number 44065/31/2020. I thank my colleague Mikko Harjanne for useful discussions and support, and my colleagues Fein Sun and Markku Kapulainen for respectively fabricating and measuring the devices presented in §5.